\documentclass{svproc}
\usepackage{url}

\usepackage{cite}
\usepackage{amsmath,amssymb,amsfonts}
\usepackage{algorithmic}
\usepackage{graphicx}
\usepackage{textcomp}

\usepackage{pifont}
\usepackage{adjustbox}

\usepackage{color, colortbl}

\usepackage{multirow}
\usepackage[utf8]{inputenc}
\usepackage[table]{xcolor}

\begin{document}
\mainmatter              
\title{Modeling Traffic Congestion in Developing Countries using Google Maps Data}
\titlerunning{Congestion Modeling using Google Maps Data}  

\author{Md. Aktaruzzaman Pramanik\inst{1} \and Md Mahbubur Rahman \inst{2} \and
ASM Iftekhar Anam\inst{3} \and Amin Ahsan Ali\inst{1} \and M Ashraful Amin\inst{1} \and A K M Mahbubur Rahman\inst{1}}

\authorrunning{Md. Aktaruzzaman Pramanik et al.}
\institute{Independent University, Dhaka, Bangladesh, \\
\email{ a.pramanikk@gmail.com, aminali@iub.edu.bd,  aminmdashraful@iub.edu.bd, akmmrahman@iub.edu.bd}
\and 
Crowd Realty	,
\email{mahbuburrahman2111@gmail.com}
\and
University of Wisconsin - Green Bay, Green Bay, USA, 
\email{anami@uwgb.edu}}

\maketitle              

\begin{abstract}Traffic congestion research is on the rise,  thanks to urbanization, economic growth, and industrialization. Developed countries invest a lot of research money in collecting traffic data using Radio Frequency Identification (RFID), loop detectors, speed sensors, high-end traffic light, and GPS. However, these processes are expensive, infeasible, and non-scalable for developing countries with numerous non-motorized vehicles, proliferated ride-sharing services, and frequent pedestrians. This paper proposes a novel approach to collect traffic data from Google Map's traffic layer with minimal cost. We have implemented widely used models such as Historical Averages (HA), Support Vector Regression (SVR), Support Vector Regression with Graph (SVR-Graph), Auto-Regressive Integrated Moving Average (ARIMA) to show the efficacy of the collected traffic data in forecasting future congestion. We show that even with these simple models, we could predict the traffic congestion ahead of time. We also demonstrate that the traffic patterns are significantly different between weekdays and weekends. 

\end{abstract}

\section{Introduction}
Traffic congestion has been a growing concern for cities around the world. Rapid urbanization, limited space for expansion, and increasing demand for transportation are turning the congestion more vulnerable in the cities, especially in developing countries. Along with the factors mentioned above, the widespread availability of ride-sharing services and non-motorized vehicles are also responsible for the overcrowded streets in those cities. Traffic congestion affects the lives of the people living in urban areas; specifically, it increases mental stress and disrupts people's daily schedules, resulting in elevated blood pressure, increased negative mood states, and lowered tolerance for frustration. Moreover, traffic congestion has negative impacts on the economy. It increases business production costs due to longer travel times, missed deliveries, and increased fuel costs \cite{falcocchio2015costs}.

In recent years, transportation research focuses on traffic congestion control and minimization of congestion time,  mostly to resolve the above problems. Developed countries invest in intelligent transportation research where researchers collect traffic data through loop detectors, Radio Frequency Identification (RFID), and sensor networks to model and predict the traffic pattern \cite{chen2019review}. Such models can play a pivotal role in developing congestion prediction applications for commuters, travelers, and traffic management authorities.  

The authors of \cite{yu2017spatio} have collected traffic speed data using 39000 loop detector sensors from 12 roads in China (BJER4 dataset) with five-minute intervals. They have also used the PeMSD7 dataset \cite{chen2001freeway} obtained from  California Traffic Department to test their proposed traffic model. Another research group collected the vehicle speeds with five-minute intervals in Los Angeles County and California with 532 speed sensors  ~\cite{chen2019gated}. Guo et al. have developed traffic models based on Dataset-I, Dataset-II, and PeMSD4 \cite{chen2001freeway} that were collected from Washington DC (370 roads), Philadelphia (397 roads), and San Francisco Bay area (307 roads), respectively, using many speed sensors\cite{guo2020optimized}. The authors of ~\cite{zhang2019trafficgan} accumulated traffic data from Chicago's 1250 arterial streets via the vehicles' GPS. 
Refer to Table \ref{WWDataTable} for a comparative summary of the studies discussed above. 

\begin{table*}
\caption{Traffic data collection scenarios in developed countries}

\label{WWDataTable}
 \begin{adjustbox}{width=\textwidth}
\begin{tabular}{|c|c|c|c|c|c|c|}
\hline
\textbf{Research} & \begin{tabular}[c]{@{}c@{}}\textbf{Dataset} \\ \textbf{Used} \end{tabular} & \begin{tabular}[c]{@{}c@{}} \textbf{Location} \\ \textbf{(Road Seg.)} \end{tabular} & \begin{tabular}[c]{@{}c@{}} \textbf{Collection} \\ \textbf{Process} \end{tabular} & \begin{tabular}[c]{@{}c@{}} \textbf{Collection} \\ \textbf{Period}  \end{tabular}& \textbf{Interval}& \textbf{Limitations} \\ \hline

\begin{tabular}[c]{@{}c@{}} Yu et al.\\~\cite{yu2017spatio} \end{tabular}& \begin{tabular}[c]{@{}c@{}}BJER4 \\ PeMSD7 ~\cite{chen2001freeway}\end{tabular} & \begin{tabular}[c]{@{}c@{}}Beijing (12)\\ California\end{tabular} & 
\begin{tabular}[c]{@{}c@{}}Loop detectors\\ Sensors \end{tabular}
&\begin{tabular}[c]{@{}c@{}}1 July-31 August,\\2014, No weekends\end{tabular} &5 min& \begin{tabular}[c]{@{}c@{}}39000 \\ detectors,\\
expensive
\end{tabular} \\ \hline

\begin{tabular}[c]{@{}c@{}} Chen et al. \\~\cite{chen2019gated}  \end{tabular}& \begin{tabular}[c]{@{}c@{}}METR-LA \\ PEMS-BAY\end{tabular} & \begin{tabular}[c]{@{}c@{}}Los Angeles, \\ California\end{tabular} & 
\begin{tabular}[c]{@{}c@{}}Sensors  \end{tabular}
&\begin{tabular}[c]{@{}c@{}}1 May-30 June,2012\\
	1 Jan-31 May,2017\end{tabular}&5 min& \begin{tabular}[c]{@{}c@{}}Total 532 Sensors\\
expensive, \\not scalable
\end{tabular} \\ \hline

\begin{tabular}[c]{@{}c@{}}Guo  et al.\\~\cite{guo2020optimized} \end{tabular} & \begin{tabular}[c]{@{}c@{}}Dataset-I \\ Dataset-II\\PeMSD4~\cite{chen2001freeway}\end{tabular} & \begin{tabular}[c]{@{}c@{}}Wash.  D.C.(370)\\ Philadelphia(397)\\SF. Bay Area(307)\end{tabular} & 
\begin{tabular}[c]{@{}c@{}} Sensors \end{tabular}
&\begin{tabular}[c]{@{}c@{}}24000 measurements\\24000 measurements\\Jan-Feb 2018,\\ 15000 measurements
\end{tabular}& 5 min &  \begin{tabular}[c]{@{}c@{}}Enormous \\amount \\of sensors
\end{tabular} \\ \hline

\begin{tabular}[c]{@{}c@{}}Zhang et al.\\~\cite{zhang2019trafficgan} \end{tabular} & \begin{tabular}[c]{@{}c@{}}CTA Data ~\cite{CTAdata} \end{tabular} & \begin{tabular}[c]{@{}c@{}}Chicago’s \\streets(1250)\end{tabular} & 
\begin{tabular}[c]{@{}c@{}} GPS Tracing \end{tabular}
 &\begin{tabular}[c]{@{}c@{}}6 months \end{tabular}&10 min &  \begin{tabular}[c]{@{}c@{}} Infeasible for \\Non-motorized \\ vehicles \\ in developing \\countries
\end{tabular} \\ \hline

\end{tabular}
 \end{adjustbox}

\end{table*}


It is easy to notice that the above works primarily focused on the developed countries where the transportation research division has a substantial budget to collect traffic data using loop detectors, sensors, and cameras.  Deploying such an infrastructure could be prohibitively expensive for developing countries like Bangladesh, India, or Pakistan. Moreover, these techniques are not quite effective in the cities of developing countries with the narrow streets, the prevalence of non-motorized vehicles, and non-grid-like road structure.  Some research studies attempt to collect traffic data in several developing countries, e.g., Bangladesh, Pakistan, Myanmar, and India. 
S.M. Labib et al. collected data with manual counts by deploying local surveyors for their case study area: Sheraton hotel junction in Dhaka, Bangladesh \cite{labib2019integrating}.  Another group collected traffic data by manually counting the vehicles at Science Laboratory Intersection in peak periods of the morning (9:00 -- 10:00 AM)~\cite{roy2015study}.

Some researchers collected traffic data using GPS for different purposes, such as GPS data from several taxis for 11769 road segments ~\cite{rahman2018traffic}. Salma et al.   collected  GPS  data from a  telecommunication company of Bangladesh with vehicles with GPS tracking devices\cite{salma2018enhancing}. However, they have collected traffic information with an interval of one hour.  The interval is too high, and the number of vehicles is limited for modeling the traffic patterns.

\begin{table*}[!t]
\caption{Traffic Data Collection Scenarios in Some South Asian Developing Countries}
\label{BDTable}
 \begin{adjustbox}{width=\textwidth}
\begin{tabular}{|c|c|c|c|c|c|c|}
\hline
\textbf{Research} & \begin{tabular}[c]{@{}c@{}} \textbf{Location} \\  \textbf{(Coverage)} \end{tabular} & \begin{tabular}[c]{@{}c@{}} \textbf{Collection} \\  \textbf{ Process} \end{tabular} & \begin{tabular}[c]{@{}c@{}} \textbf{Collection} \\ \textbf{ Period} \end{tabular} &\textbf{Interval}  &\textbf{Goal} & \textbf{Limitations} \\ \hline

\begin{tabular}[c]{@{}c@{}}Labib et al.\\~\cite{labib2019integrating} \end{tabular}& \begin{tabular}[c]{@{}c@{}}Sheraton Hotel \\ Junction,\\Dhaka\\(1 intersection)\end{tabular} & 
\begin{tabular}[c]{@{}c@{}} Manual counts \end{tabular}
& \begin{tabular}[c]{@{}c@{}}7:00 to 11:00 AM \end{tabular} & 15 mins & \begin{tabular}[c]{@{}c@{}}
Optimization \\ of signal timing 
\end{tabular} &\begin{tabular}[c]{@{}c@{}} \\Erroneous counting,\\ non-scalable,\\time consuming
\end{tabular} \\ \hline

\begin{tabular}[c]{@{}c@{}}Roy et al.\\~\cite{roy2015study} \end{tabular}& \begin{tabular}[c]{@{}c@{}}Science lab\\to Elephant road\\intersection, Dhaka\\(1 Intersection)\end{tabular}  & 
\begin{tabular}[c]{@{}c@{}}Manual counts \end{tabular}
& \begin{tabular}[c]{@{}c@{}}9:00 to 10:00 AM \end{tabular} & 15 mins & \begin{tabular}[c]{@{}c@{}}
VISSIM\\simulations
\end{tabular} &\begin{tabular}[c]{@{}c@{}}Erroneous  \\and \\cumbersome \\manual counting 
\end{tabular} \\ \hline

\begin{tabular}[c]{@{}c@{}}Rahman et al.\\~\cite{rahman2018traffic}\end{tabular} & \begin{tabular}[c]{@{}c@{}}11,769 road segments \\Dhaka\end{tabular}  & 
\begin{tabular}[c]{@{}c@{}}GPS Data \end{tabular}
& \begin{tabular}[c]{@{}c@{}}N/A \end{tabular} & 30 mins & \begin{tabular}[c]{@{}c@{}}
Learn traffic\\ patterns
\end{tabular} & \begin{tabular}[c]{@{}c@{}}Not usable for \\traffic intensity\\prediction
\end{tabular} \\ \hline

\begin{tabular}[c]{@{}c@{}}Salma et al. \\ ~\cite{salma2018enhancing}\end{tabular} & \begin{tabular}[c]{@{}c@{}}Airport \\ to Gulshan-2,\\ Dhaka\\(9048 meter)\end{tabular}  & 
\begin{tabular}[c]{@{}c@{}}GPS Data,\\ Telecom \\ company vehicle \end{tabular}
& \begin{tabular}[c]{@{}c@{}}working day \\(peak hour),\\weekend, \\Last day  of  week. \end{tabular} & 1 hour & \begin{tabular}[c]{@{}c@{}}
Predicting \\traffic intensity
\end{tabular} & \begin{tabular}[c]{@{}c@{}}No traffic  \\intensity data,  \\Interval is too high
\end{tabular} \\ \hline

\begin{tabular}[c]{@{}c@{}}Lwin et al.\\~\cite{lwin2015estimation} \end{tabular}& \begin{tabular}[c]{@{}c@{}}Yangon,\\ Myanmar\end{tabular} & \begin{tabular}[c]{@{}c@{}}GPS data,\\mobile phones \\on vehicle \end{tabular} 
& \begin{tabular}[c]{@{}c@{}}4 months \\of GPS data \\ \end{tabular} & N/A &\begin{tabular}[c]{@{}c@{}}Probability of user's\\ source \\ and\\ destination \end{tabular} & \begin{tabular}[c]{@{}c@{}}Limited \\ to Android \\ only
\end{tabular} \\ \hline

\begin{tabular}[c]{@{}c@{}}Sharma et al.\\~\cite{sharma2018ann} \end{tabular}& \begin{tabular}[c]{@{}c@{}}Roorkee,\\Haridwar,\\Delhi,\\and M.nagar\\India \\(3 Intersections)\end{tabular}  & 
\begin{tabular}[c]{@{}c@{}} Digital \\cameras,  \\manual traffic \\volume count \end{tabular}
& \begin{tabular}[c]{@{}c@{}}9:00am to 12:00am\\
3:00pm to 6:00pm
 \end{tabular} & 5 mins&\begin{tabular}[c]{@{}c@{}}Short-term \\ traffic\\ forecasting \end{tabular} & \begin{tabular}[c]{@{}c@{}}Limited to \\ small portion of\\ highways 

\end{tabular} \\ \hline

\begin{tabular}[c]{@{}c@{}}Ali et al.\\~\cite{ali2014estimation} \end{tabular}& \begin{tabular}[c]{@{}c@{}}Karachi, \\Pakistan\\(9 locations)\end{tabular}  & 
\begin{tabular}[c]{@{}c@{}}Video camera,\\ tracker,\\ manual\\calculation \end{tabular}
& \begin{tabular}[c]{@{}c@{}}8:00am to 1:00pm\\3:00pm to 8:00pm \end{tabular} & N/A & \begin{tabular}[c]{@{}c@{}}Estimate  traffic\\congestion cost \end{tabular} & \begin{tabular}[c]{@{}c@{}}Manual \\ calculation of \\traffic delay

\end{tabular} \\ \hline

\end{tabular}

 \end{adjustbox}
\end{table*}


Several studies focused on other developing countries of the South Asian region. Ali et al. collected traffic data of two peak periods in a day of nine different locations of Karachi, Pakistan\cite{ali2014estimation}. They calculated the traffic delay manually based on the traffic volume at those locations. Another work collected GPS data for traffic congestion prediction \cite{lwin2015estimation}. For predicting the congestion on user's demand, the authors collected traffic speed, direction, timestamps,  and other GPS  data by tracing mobile phones on some predefined vehicles at Yangon, Myanmar. However, they could not perform large scale data collection as they focused on only android phones. In India, Sharma and his group used manual counting on vehicles from high-quality digital cameras mounted over a couple of two- and four-lane highways during two peak sessions of the working days\cite{sharma2018ann}. They collected data from cities Roorkee, Haridwar, Delhi, and Muzaffarnagar. Then they used machine learning models to forecast traffic congestion. Please see Table  \ref{BDTable} for the comprehensive summary. The table suggests that most of the data collection in developing countries did not target traffic congestion prediction, and the researchers relied mainly on manual counting or GPS data.

To summarize, the cited traditional data collection processes are tedious, require considerable human effort and instruments. Moreover, the approaches are not scalable for large collection scenarios for the whole city traffic. For developing countries, the traffic data needs to be collected in a configurable and scalable manner with minimal cost. In this paper, we propose to use traffic information from Google Map Service in a unique way so that the traffic data for any road segments/intersections can be collected seamlessly from any location in a city. 

At present, the Google map provides the current situation of traffic congestion. We proposed an easy way of collecting traffic data. By extracting Google Map data, we can analyze the traffic condition for selected intersections. As Google Maps itself do not provide traffic data, we have collected the traffic condition of each road segment by observing their traffic color: green (no traffic), orange (moderate traffic), red(heavy traffic), and dark red (very heavy traffic). Using the Google Traffic Layer API, we have collected this information with a 30-second interval for six months: November 2019 - April 2020. By analyzing the data from those intersections, we can develop a statistical traffic congestion model.  Our data collection process is less expensive and more efficient compared to other approaches. Moreover, it is scalable to the entire city.

In Table \ref{comparisonTable}, we have summarized the characteristics of the datasets collected by other research papers and our proposed one in terms of covered area, cost of deployment, scalability, interval flexibility, and remote data collection facility. It is easy to note that our proposed approach achieves most of the desired characteristics suitable for deployment in developing countries.

\begin{table*}[!t]
\caption{Comparison of Our Dataset with Developing Countries}
\label{comparisonTable}
 \begin{adjustbox}{width=\textwidth}
\setlength{\tabcolsep}{4pt}
\renewcommand{\arraystretch}{1}
\begin{tabular}{|c|c|c|c|c|c|}
\hline
Researches & 
\multicolumn{1}{|p{2cm}|}{\centering Larger Area\\Covering} & 
Low Cost &
Scalability &
\multicolumn{1}{|p{2cm}|}{\centering Flexibility in \\Adusting Interval}&

\multicolumn{1}{|p{2cm}|}{\centering Remote Data\\Collection} \\ \hline
Labib et al.~\cite{labib2019integrating} & 
\ding{55} & \ding{51} & \ding{55} & \ding{51}  & \ding{55} \\  \hline

Roy et al.~\cite{roy2015study} & 
\ding{55} & \ding{51} & \ding{55} & \ding{51}  & \ding{55} \\  \hline

Rahman et al.~\cite{rahman2018traffic} & 
\ding{51} & \ding{51} & \ding{55} & \ding{55}  & \ding{51} \\  \hline

Salma et al.~\cite{salma2018enhancing} & 
\ding{55} & \ding{51} & \ding{55} & \ding{55}  & \ding{55} \\  \hline

Lwin et al.~\cite{lwin2015estimation} & 
\ding{55} & \ding{55} & \ding{51} & \ding{55}  & \ding{51} \\  \hline

Sharma et al.~\cite{sharma2018ann} & 
\ding{55} & \ding{55} & \ding{55} & \ding{51} & \ding{55} \\ \hline

Ali et al.~\cite{ali2014estimation} & 
\ding{51} & \ding{55} & \ding{55} & \ding{51}  & \ding{55} \\  \hline

\textbf{Our Proposed Approach} & 
\ding{51} & \ding{51} & \ding{51} & \ding{51}  & \ding{51} \\ \hline
\end{tabular}
 \end{adjustbox}

\end{table*}

Finally, we have developed some well-known statistical (Historical average, ARIMA) and machine learning (Support Vector Regression (SVR), SVR with graph) predictive models to show the efficacy of the collected data. We used one month's data  (November 2019) from Mirpur - Dhaka in our analysis. Therefore, our contribution includes:
\begin{itemize}
    \item We propose a novel data collection method for traffic data collection that is scalable, low cost, and efficient for developing countries
    \item We have demonstrated the efficacy of our traffic data in predicting traffic congestion in all intersections of the Mirpur area. 
    \item We have shown the comparative analysis of how the history of traffic congestion length affects a predictive model's capability.
    
    \item Finally, we have shown the significant difference between traffic congestion patterns between weekdays and weekends in Mirpur, Dhaka
\end{itemize}

The remainder of this paper is organized as follows:
Section \ref{related-works} describes the related literature about traffic data collection and forecasting. Section \ref{data-collection} introduces the proposed traffic data collection. Section \ref{statistical-machine-learning-models} defines the statistical and machine  learning models for traffic prediction. Section \ref{experiments-results} presents the experimental setup and performance comparisons between different prediction models. The results also include performance analysis of traffic prediction in different time duration during weekdays and weekends. Finally, Section \ref{conclusion} concludes the paper and discusses future studies.

\section{Related Works}
\label{related-works}
Much research has been done in recent years on traffic congestion prediction and traffic forecasting based on time-series data \cite{faloutsos2019classical}, \cite{kuznetsov2018foundations}. In the case of developing countries, the variation of motorized and non-motorized vehicles is enormous \cite{labib2014transport}. For various reasons, the traffic-congestion in several areas has increased, the increasing traffic jam has become a threat to the major metropolitan cities, and its impact is observable \cite{mahmud2012possible}.
In several works, traffic data was collected using GPS tracing or using the sensors located at several intersections ~\cite{zambrano2018modeling}, \cite{iyer2018urban}, \cite{liu2018urban},  \cite{akbar2017predictive}. 

Alghamdi and colleagues applied Auto-Regressive Integrated Moving Average(ARIMA)-based modeling in their short-term time series data with 13 attributes collected from traffic flow in four different lanes \cite{alghamdi2019forecasting}. They have tuned different ARIMA parameters for better accuracy but did not show any comparison with other models. Iyer et al. proposed Deep Learning architecture(LSTM) for the prediction of congestion state of road segments using their dataset, which was collected by GPS tracing the specific vehicles ~\cite{iyer2018urban}. Liu et al. applied Deep Learning models(CNN, RNN, SAE) on their mobility data collected by both some intelligent devices (e.g., loop detectors, traffic cameras) and a large number of public vehicles equipped with GPS devices ~\cite{liu2018urban}. Their data collection methods are quite expensive and complicated compared to the metropolitan cities in developing countries where vehicles have a considerable variety and density. 

Zambrano-Martinez and colleagues applied logistic regression to characterize travel times and classified several street segments into three categories based on the measured number of vehicles from their Simulation of Urban MObility (SUMO \cite{behrisch2011sumo}) simulations \cite{zambrano2018modeling}.  In \cite{zhong2018new}, the authors proposed a neural network prediction model based on LSTM to predict traffic congestion at Shenzhen, China. The data set is collected by the traffic camera at all crossroads. This data collection process is expensive, and its almost impossible to implement in developing countries, i.e., Dhaka, Bangladesh. Research suggests that historical sensor-based data can provide less accuracy while predicting traffic state. For that reason, Ahmed and his group proposed the CTM-EKF-GA framework in which they did not use any historical traffic data; instead, they collected data from connected vehicles(CVs), which is considered real-time traffic data ~\cite{ahmed2019real}. It is also an expensive way of collecting traffic data.

Yao et al.  conducted their experiments on real datasets, such as taxi and bike trips records in New York City ~\cite{yao2018modeling}. Collecting this type of data is impossible in cities like Dhaka, Karachi, or Delhi, as every route contains varieties of motorized and non-motorized vehicles, and they are equally responsible for traffic congestion. Some groups collected real-world datasets of two types using many sensors/loop detectors in Los Angeles County, California, Beijing City, and District 7 of California \cite{li2017diffusion}, \cite{yu2017spatio}, \cite{guo2019attention}, \cite{chen2019gated}. However, collecting large-scale data by setting up expensive loop detectors in highways is hardly feasible in developing countries. While in~\cite{zheng2016urban}, they have shown an approach of using inexpensive Big Data from buildings in Hong Kong.  By using building population data, they tried to predict the traffic density in the nearby area. They conducted a correlation between the ICC(Hong Kong International Commerce Centre) occupancy data and the traffic data for the roads of different distances from the ICC. Cui et al. have collected two real-world datasets; the first one contains traffic state data collected from 323 sensor stations in the Greater Seattle Area~\cite{cui2019traffic}. The second dataset contains road link-level traffic speeds aggregated from GPS probe data collected by commercial vehicle fleets and mobile apps provided by INRIX. While ~\cite{zhang2019trafficgan} have collected their dataset from Chicago Transit Authority(CTA) buses on Chicago's arterial streets in real-time by GPS traces, no organization in Bangladesh provides this kind of traffic data collected by GPS traces for any particular area.

There is a plethora of traffic research that has been done in developing countries. They have either done traffic simulation with vehicle counts or forecasting of traffic speed from GPS data. In ~\cite{labib2019integrating}, they have collected traffic data by manual counts employing local experts in several junctions and used traffic simulation at the micro level to create optimized signal timings for a congested intersection. Roy et al. collected traffic volume data manually in peak periods of the morning(9:00–10:00 AM) and collected geometric and control data for four intersections of Dhaka city and simulated the data using VISSIM optimization of existing traffic signals \cite{roy2015study}. For identifying the traffic density patterns of Dhaka across different roads, ~\cite{rahman2018traffic} used GPS data for 11769 road segments.  Furthermore,  In \cite{salma2018enhancing}, another group collected GPS data from a telecommunication company of Bangladesh who has vehicles with GPS  tracking devices. The authors have accumulated traffic information with an interval of one hour, which is very high compared to the state of the art traffic data intervals because of traffic status changes in smaller intervals (Table \ref{BDTable}).

For predicting traffic congestion probability on user's demand, \cite{lwin2015estimation} collected traffic speed, direction, timestamps, and other GPS data at Yangon, Myanmar, by tracing mobile phones on vehicles. In ~\cite{sharma2018ann}, the authors collected traffic data by manually counting vehicles on high-quality digital cameras, two or four-lane highway stretches of India. They have built a short-term traffic forecasting model by using the collected data. From the works mentioned in this section, we have identified some limitations in developing countries. In summary, they are
\begin{itemize}
    \item Deploying sensors and loop detectors that are expensive.
    \item Collecting GPS data from a representative number of motorized vehicles, which is not feasible.
    \item Presence of a significant amount of non-motorized vehicles in the city streets.
    \item Manual counting is tedious, time-consuming, and non-scalable.
    \item Collecting traffic data for the whole city is practically impossible.
\end{itemize}

In the next section, we have described our proposed data collection process that overcomes the aforementioned limitations.

\begin{figure}[!t]
\centering
\includegraphics{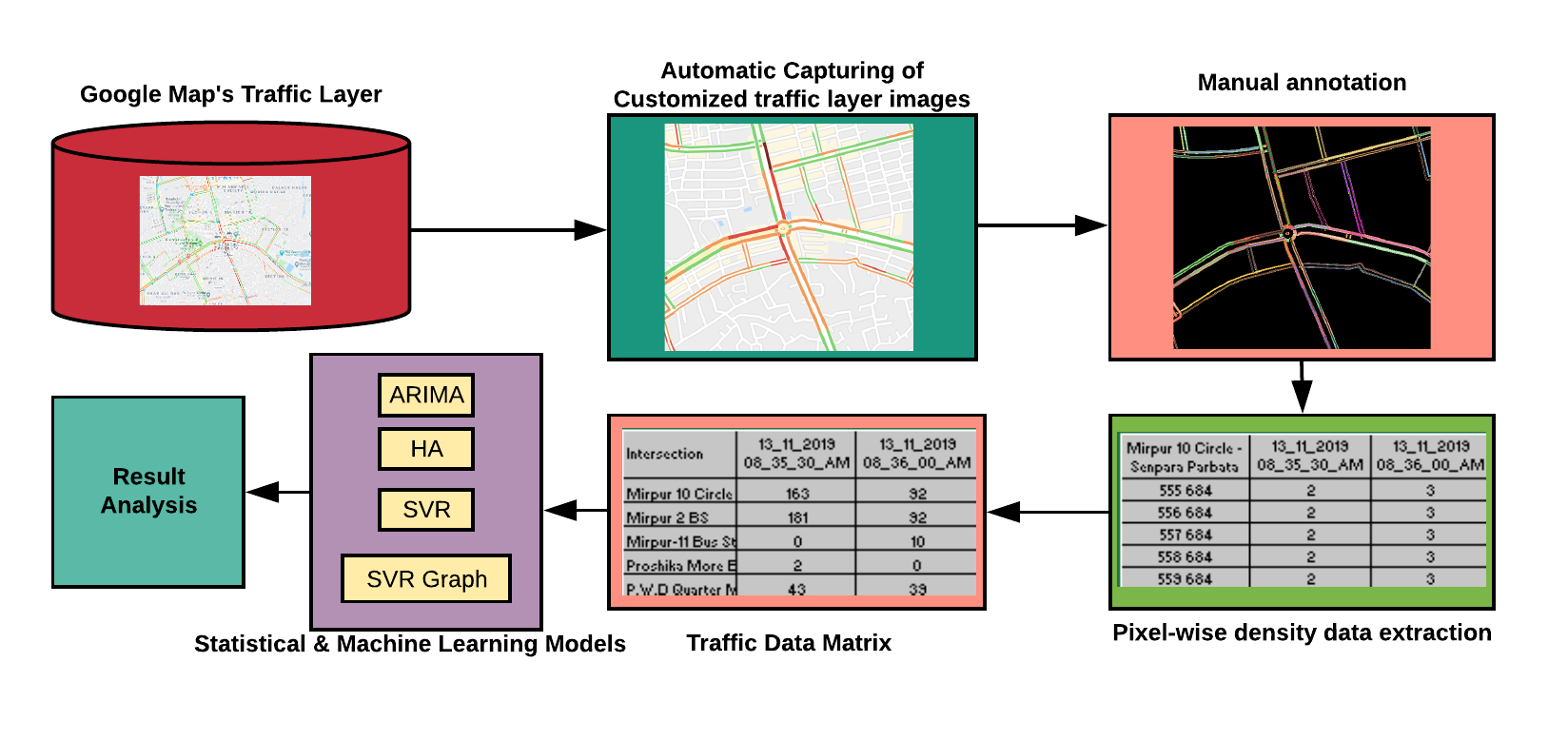}
\label{blockDiagram}
\caption{Block Diagram of the proposed data collection and prediction}
\end{figure}

\begin{figure}[!tbhp]
\centering
\includegraphics[scale=0.9]{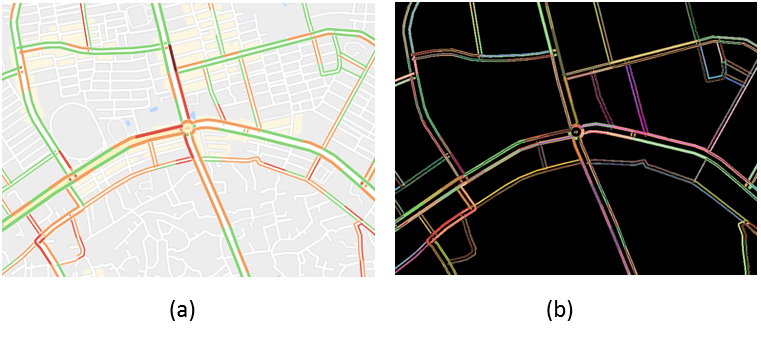}
\label{gmap_to_cmap}
\caption{Sample output of data capture (a)  and preprocessing (b)}
\end{figure}

\begin{figure}[!t]
\centering
\includegraphics[scale=0.9]{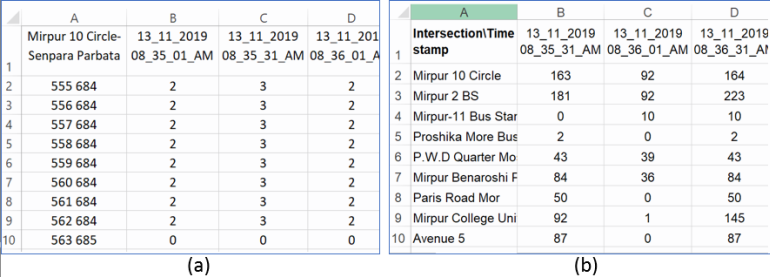}
\label{pixel_to_matrix}
\caption{Sample data after  of image processing (a)  and Final data matrix for predictive model (b)}
\end{figure}

\section{Data Collection}
\label{data-collection}

Figure 1 
shows the steps in our proposed data collection and prediction framework. The following subsections describe each of the steps in more detail. 

\subsection{Automatic Google data capture}
We have collected Google Map's traffic layer images of Mirpur, Dhaka (latitude: 23.8060493, longitude: 90.3712275) with a zoom level of 15 on Google Maps covering approximately 12.561 \(km^2\) area. Google does not provide traffic layer images of a specified time interval. So we developed a tool to capture traffic layer images automatically. We have used Google Maps' JavaScript API to customize the image, such as removing the labels of bus-stops, roads, transit, and administrative neighborhoods from the map. We need to provide three parameters for our data collection program: location(latitude-longitude), zoom level on Google Maps, and time interval for capturing each image. With these inputs, the tool automatically captures and saves images. Figure \ref{gmap_to_cmap}(a) shows a sample map image. For our selected area, we have captured images from 6:00 AM to 11:59 PM with an interval of 30 seconds, which stands 2160 images a day and 64800 images for a month. We have collected images of 6 months, from November 2019 to April 2020. We have used an interval that is frequent enough to capture the traffic changes and does not cross the daily free quota of the Google Maps API.

\subsection{Image prepossessing}
After capturing the images, we prepared them to extract traffic density data. We annotated each road segment in one image and assigned an ID and a color code (RGB) to each road segment. In this process, we have identified 155 road segments and 45 intersections in the Mirpur area. Figure \ref{gmap_to_cmap}(b) shows the color-annotated map image. Next, we extracted pixel locations for each road segment using their assigned RGB values. We adopted OpenStreetMap standard annotation to assign the IDs to each intersection. Thus, it can be extended to any arbitrary location. 

\subsection{Traffic data extraction}
Using the pixel locations collected in the previous step, we extracted each road segment's pixel value. This time, we have pixel-wise traffic density information for each road segment(both incoming and outgoing) of an intersection. The numeric values 1, 2, 3, 4, and 0 mean no traffic delay, medium amount of traffic, near-heavy traffic, heavy traffic, and no traffic information. This information is assigned in this format because Google Map's traffic layer shows four colors(green, orange, red and dark red) to represent the amount of traffic density of a road segment\footnote{https://support.google.com/maps/answer/3092439?co=GENIE.Platform}.

\subsection{Data processing}
In the final step of our data processing, we took every intersection and summed the intersection's incoming traffic density. We have only considered the traffic density values 3 and 4, which indicate near-heavy and heavy traffic density. As we know, incoming traffic is mainly responsible for creating traffic congestion in an intersection, so that, for a specific timestamp, we have used the summation of incoming traffic as a measure of traffic density of a particular intersection. Figure 3 
shows the image data transformed into matrices for predictive modeling.

\section{Statistical and Machine Learning models for predictions}
\label{statistical-machine-learning-models}

\subsection{Historical Average Forecast}
Traffic flow in a particular area may follow some patterns because of some regular events occurring there. For that reason, we may expect that historical averages of traffic flow or density of a particular area at a particular time should provide us a significant traffic density forecast of that area at the same time and on the same weekday.

\subsection{Support Vector Regression(SVR)}
We have also used the Support Vector Regression(SVR) model for traffic density prediction. In Support Vector Regression, the training dataset includes multivariate sets of observations and observed response values.\\
While in linear SVR, if \(x_i\) is the \(i^{th}\) traffic intensity  of N observations with observed response traffic intensity values \(y_i\) and if we use
nonnegative multipliers \(\alpha_i\) and \(\alpha_i^*\) for each observation \(x_i\), the prediction function becomes-

\begin{equation}
    y = \sum_{i=1}^{N} (\alpha_i - \alpha_i^*)\cdot(x_i,x)+b
\end{equation}

In the case of our traffic density dataset, we can not predict using a linear model, that is why we replace the dot product \(x_i,x_j\) with a nonlinear kernel function \(K(x_i,x_j)=exp(-\frac{||x_i-x_j||^2}{2\sigma^2})\), which transforms the data into higher dimensional feature space and helps to perform the linear separation. So the equation becomes-    
\begin{equation}
    y = \sum_{i=1}^{N} (\alpha_i - \alpha_i^*)\cdot K(x_i,x)+b
\end{equation}

\subsection{Graph SVR}
Sometimes, the traffic congestion in a particular intersection is highly influenced by the amount of traffic present in the neighboring intersections. We have used the  Support Vector Regression(SVR) model in a different approach where we have considered the summation of the traffic density of neighboring nodes for each observation \(x_i\). So this time, the SVR model predicts the future traffic density for a particular intersection considering the certain traffic status of neighboring intersections. 
\subsection{Autoregressive Integrated Moving Average (ARIMA)}
ARIMA (also known as the Box-Jenkins model) is a simple but powerful model that we have used to train our time-series traffic density data for forecasting future traffic congestion.  ARIMA model consists of autoregressive terms(AR), moving average terms(MA), and differencing operations(I).

Before applying ARIMA model, we must need to conform to the stationarity of the dataset. A degree of differencing
reduces non-stationarity (seasonality) from the dataset. Such as,  d=1 means first-order differencing \(Z_t = X_t - X_{t-1}\), again, when differencing order = 2, \(z_t = (x_t - x_{t-1}) - (x_{t-1} - x_{t-2})\). In ARIMA(p,d,q)
model the parameter \texttt{d} defines the differencing order.
The parameter \texttt{p}  defines the number of prior observations that have a significant correlation with the current observation. While \texttt{q} represents the moving window size in terms of error, which impacts current observation. The ARMA(p,q) equation-

\begin{multline}
     Y_t = \Phi_1Y_{t-1}+\Phi_2 Y_{t-2}+...+\Phi_3 Y_{t-p}+\\
    +\omega_1\epsilon_{t-1}+\omega_2\epsilon_{t-2}+...+\omega_q\epsilon_{t-1}+\epsilon_t
\end{multline}

where the weights \((\Phi_1,\Phi_2...\Phi_3)\) and 
\((\omega_1,\omega_2...\omega_3)\) respectively for the AR and MA are calculated depending on the correlations between the lagged observations and current observation. For our dataset, we have chosen ARIMA(1,0,0).

\section{Experiments and Results}
\label{experiments-results}
We performed some experiments to show the efficacy of the collected traffic data in predicting congestion intensity.  Before going to the experimental setups, we would like to introduce various hyper-parameters: sampling rate, sequence length, and prediction length that impact the performance.  \\
\textbf{Sampling rate} denotes the time interval at which traffic intensities are used for each node. \\
\textbf{Sequence length} represents the length of the input data across the previous timestamps from the current time. The sequence data is necessary to predict the traffic intensity in the future.  \\
\textbf{Prediction length} indicates the number of future timestamps when the models would predict traffic intensities.
\subsection{Experimental Setup}
\subsubsection{Prediction of Traffic Intensity}
In this experiment, we have used 20 days of traffic intensity data as training and ten days of data for testing. In these experiments, we have varied the hyper-parameters: sampling rate, sequence length, and prediction length to find what setting yields the best prediction results.
\subsubsection{Traffic Analysis in Weekdays}
Since the traffic patterns of Dhaka significantly vary on weekdays and weekends, we have performed an experiment that has taken these scenarios into account. In November 2019, there were 20 weekdays. We have trained the models with 14 days and tested using six days. In this experiment, we have used the hyper-parameters that yielded the best results in the earlier experiments.

\subsubsection{Traffic Prediction: Weekdays vs. Weekends}  In order to show the difference between traffic patterns on weekdays and weekends, we have performed a couple of experiments. In one experiment, we have taken 20 weekdays data as the training set and ten weekend days data as the test set. In the subsequent experiment, we have chosen seven days of weekend data as training and the last three weekday data as test data.
\subsubsection{Traffic Analysis in Weekends}
To analyze the model's performance during the weekends, we have designed this experiment where both training and test data are chosen from the weekends. In November 2019, we had a total of 10 days of weekends. Among them, we have chosen seven days of train and three days of tests.

\subsection{Performance Metrics} Since the ground truth and predicted traffic intensities are all real numbers, we have used the following metrics to compare different statistical and machine learning models.
\subsubsection{Root Mean Square Error (RMSE)} This metric is calculated by taking the root mean square of the ground truth vector and the predicted vector by the following equation

\begin{equation}
      RMSE = \sqrt{\sum_{i=1}^{N}\frac{(\hat{y}_i-y_i)^2}{N}}
\end{equation}

Here,  $y_i$ is the ground truth, and $\hat{y}_i$ is the predicted traffic intensity for  $i^{th}$ data sample. The lower value of RMSE represents better prediction performance.

\subsubsection{Mean Absolute Error} This measure has been calculated by finding the mean absolute error between ground truth and predicted traffic intensity vector. The equation is given below:
\begin{equation}
    MAE = \frac{1}{N}\sum_{i=1}^{N}|y_i-\hat{y}_i|
\end{equation}

MAE's lower value is desirable because it indicates that the ground truth and prediction are close to each other.

\subsubsection{Correlation Coefficient ($CORR$)}
We have used Pearson's Coefficient of Correlation for calculating the relationship between our ground truth and predicted traffic intensity values. The equation is

\begin{equation}
    CORR = \frac{\sum(x-\Bar{x})(y-\Bar{y})}{\sqrt{\sum(x-\Bar{x})^2\sum(y-\Bar{y})^2}}
\end{equation}

Here x and y are respectively ground truth and predicted traffic intensity values, while
$\Bar{x}$ and $\Bar{y}$ are the mean of the ground truth and predicted traffic intensity, respectively. Higher CORR means the predicted value follows the trends of the ground truth. Therefore, high CORR is desirable in the experiments.

\subsection{Results and Discussions}
In this section, we present the results of the experiments that have been designed in the last subsection.

\begin{table*}[!thbp]
\centering
\caption{Different sample interval, sequence length, and prediction length used for the experiments}
\label{tableParameters}
\begin{tabular}{ |c|c|c|c| } 
 \hline
 Index & \begin{tabular}[c]{@{}c@{}} Sample Interval \\(Min) \end{tabular} & \begin{tabular}[c]{@{}c@{}}Sequence length\\ (Min) \end{tabular}& \begin{tabular}[c]{@{}c@{}}Prediction Length\\ (Min) \end{tabular}\\ \hline
 1 & 0.5 & 15 & 5 \\ \hline
 2 & 1 & 30 & 15 \\ \hline
 3& 5 & 45 & 30 \\  \hline
 4 & - & 60 & 45 \\  \hline
  5 & -  & - & 60 \\ 
 \hline
\end{tabular}
\end{table*}

\begin{table}[]
\caption{Eight combinations of sampling interval, sequence length, and  prediction length with top performance}
\label{performanceAll}
 \begin{adjustbox}{width=\textwidth}
\begin{tabular}{|c|c|c|c|c|c|c|}
\hline
\multicolumn{1}{|l|}  {Sample Interval} & \multicolumn{1}{l|}{Sequence Length} &  \multicolumn{1}{l|}{Predicted Length} & Models     & RMSE     & MAE      & CORR   \\ \hline

&                                                 &                                                 & HA         & \underline{\textbf{73.13}} & \underline{\textbf{42.89}} & \underline{\textbf{0.64}} \\ \cline{4-7} 
                                                    &                                                 &                                                 & SVR        & 97.04 & 72.37 & 0.52 \\ \cline{4-7} 
                                                    &                                                 &                                                 & SVR\_GRAPH & 95.82 & 72.18 & 0.52\\ \cline{4-7} 
\multirow{-4}{*}{0.5min}                            & \multirow{-4}{*}{45min}                         & \multirow{-4}{*}{5min}                          & ARIMA      & 73.18  & 42.92 & 0.64  \\ \hline
\rowcolor[HTML]{C0C0C0} 
\cellcolor[HTML]{C0C0C0}                            & \cellcolor[HTML]{C0C0C0}                        & \cellcolor[HTML]{C0C0C0}                        & HA         & {75.44} & {44.28} & {0.62} \\ \cline{4-7} 
\rowcolor[HTML]{C0C0C0} 
\cellcolor[HTML]{C0C0C0}                            & \cellcolor[HTML]{C0C0C0}                        & \cellcolor[HTML]{C0C0C0}                        & SVR        & 99.19 & 74.15 & 0.51 \\ \cline{4-7} 
\rowcolor[HTML]{C0C0C0} 
\cellcolor[HTML]{C0C0C0}                            & \cellcolor[HTML]{C0C0C0}                        & \cellcolor[HTML]{C0C0C0}                        & SVR\_GRAPH & 96.94 & 73.01 & 0.51 \\ \cline{4-7} 
\rowcolor[HTML]{C0C0C0} 
\multirow{-4}{*}{\cellcolor[HTML]{C0C0C0}0.5   min} & \multirow{-4}{*}{\cellcolor[HTML]{C0C0C0}45min} & \multirow{-4}{*}{\cellcolor[HTML]{C0C0C0}15min} & ARIMA      & 75.75 & 44.59 & 0.61 \\ \hline
                                                    &                                                 &                                                 & HA         & 78.03 & 45.89 & 0.59 \\ \cline{4-7} 
                                                    &                                                 &                                                 & SVR        & 101.15 & 75.85 & 0.50 \\ \cline{4-7} 
                                                    &                                                 &                                                 & SVR\_GRAPH & 98.14 & 73.91 & 0.50\\ \cline{4-7} 
\multirow{-4}{*}{0.5   min}                         & \multirow{-4}{*}{45min}                         & \multirow{-4}{*}{30min}                         & ARIMA      & 78.49 & 46.26 & 0.58 \\ \hline
\rowcolor[HTML]{C0C0C0} 
\cellcolor[HTML]{C0C0C0}                            & \cellcolor[HTML]{C0C0C0}                        & \cellcolor[HTML]{C0C0C0}                        & HA         & 80.37 & 47.31 & 0.57\\ \cline{4-7} 
\rowcolor[HTML]{C0C0C0} 
\cellcolor[HTML]{C0C0C0}                            & \cellcolor[HTML]{C0C0C0}                        & \cellcolor[HTML]{C0C0C0}                        & SVR        & 102.13 & 76.60 & 0.49 \\ \cline{4-7} 
\rowcolor[HTML]{C0C0C0} 
\cellcolor[HTML]{C0C0C0}                            & \cellcolor[HTML]{C0C0C0}                        & \cellcolor[HTML]{C0C0C0}                        & SVR\_GRAPH & 98.86 & 74.53 & 0.50 \\ \cline{4-7} 
\rowcolor[HTML]{C0C0C0} 
\multirow{-4}{*}{\cellcolor[HTML]{C0C0C0}0.5   min} & \multirow{-4}{*}{\cellcolor[HTML]{C0C0C0}45min} & \multirow{-4}{*}{\cellcolor[HTML]{C0C0C0}45min} & ARIMA      & 80.75 & 47.53 & 0.56 \\ \hline
                                                    &                                                 &                                                 & HA         & 73.61  & 43.02 & 0.64 \\ \cline{4-7} 
                                                    &                                                 &                                                 & SVR        & 96.93 & 72.19 & 0.53 \\ \cline{4-7} 
                                                    &                                                 &                                                 & SVR\_GRAPH & 95.03  & 71.23 & 0.53 \\ \cline{4-7} 
\multirow{-4}{*}{1.0min}                            & \multirow{-4}{*}{60min}                         & \multirow{-4}{*}{5min}                          & ARIMA      & \underline{\textbf{73.26}} & \underline{\textbf{42.70}} & \underline{\textbf{0.64}} \\ \hline
\rowcolor[HTML]{C0C0C0} 
\cellcolor[HTML]{C0C0C0}                            & \cellcolor[HTML]{C0C0C0}                        & \cellcolor[HTML]{C0C0C0}                        & HA         & {75.71} & {44.27} & {0.62}    \\ \cline{4-7} 
\rowcolor[HTML]{C0C0C0} 
\cellcolor[HTML]{C0C0C0}                            & \cellcolor[HTML]{C0C0C0}                        & \cellcolor[HTML]{C0C0C0}                        & SVR        & 99.09 & 74.25 & 0.51  \\ \cline{4-7} 
\rowcolor[HTML]{C0C0C0} 
\cellcolor[HTML]{C0C0C0}                            & \cellcolor[HTML]{C0C0C0}                        & \cellcolor[HTML]{C0C0C0}                        & SVR\_GRAPH & 96.27 & 72.20 & 0.52 \\ \cline{4-7} 
\rowcolor[HTML]{C0C0C0} 
\multirow{-4}{*}{\cellcolor[HTML]{C0C0C0}1.0min}    & \multirow{-4}{*}{\cellcolor[HTML]{C0C0C0}60min} & \multirow{-4}{*}{\cellcolor[HTML]{C0C0C0}15min} & ARIMA      & 75.93 & 44.49 & 0.61    \\ \hline
                                                    &                                                 &                                                 & HA         & 78.28 & 45.79 & 0.59  \\ \cline{4-7} 
                                                    &                                                 &                                                 & SVR        & 101.03 & 76.07 & 0.50 \\ \cline{4-7} 
                                                    &                                                 &                                                 & SVR\_GRAPH & 97.47 & 73.20 & 0.51\\ \cline{4-7} 
\multirow{-4}{*}{1.0min}                            & \multirow{-4}{*}{60min}                         & \multirow{-4}{*}{30min}                         & ARIMA      & 78.65 & 46.12 & 0.59\\ \hline
\rowcolor[HTML]{C0C0C0} 
\cellcolor[HTML]{C0C0C0}                            & \cellcolor[HTML]{C0C0C0}                        & \cellcolor[HTML]{C0C0C0}                        & HA         & 80.28 & 47.01 & 0.57\\ \cline{4-7} 
\rowcolor[HTML]{C0C0C0} 
\cellcolor[HTML]{C0C0C0}                            & \cellcolor[HTML]{C0C0C0}                        & \cellcolor[HTML]{C0C0C0}                        & SVR        & 101.77 & 76.842 & 0.49 \\ \cline{4-7} 
\rowcolor[HTML]{C0C0C0} 
\cellcolor[HTML]{C0C0C0}                            & \cellcolor[HTML]{C0C0C0}                        & \cellcolor[HTML]{C0C0C0}                        & SVR\_GRAPH & 98.19 & 73.79 & 0.50 \\ \cline{4-7} 
\rowcolor[HTML]{C0C0C0} 
\multirow{-4}{*}{\cellcolor[HTML]{C0C0C0}1.0min}    & \multirow{-4}{*}{\cellcolor[HTML]{C0C0C0}60min} & \multirow{-4}{*}{\cellcolor[HTML]{C0C0C0}45min} & ARIMA      & 80.75 & 47.31 & 0.56 \\ \hline
                                                     \hline
\end{tabular}
\end{adjustbox}
\end{table}
Table \ref{tableParameters} shows the list of parameters that we have exhaustively used in our experiments to find what combinations are the best for traffic estimation. We have used three different intervals between the samples (time interval), the sequence length of the input pattern (time length of the input traffic data), and the prediction length (the scheduled time from the current traffic where we need to predict). The interval between samples is 0.5 min (initial sampling rate: 1 sample in 30 sec), 1 min (1 sample in 1 minute), and 5 minutes (1 sample in 5 minutes). The input sequence length differs from 15 minutes to 60 minutes, where the number of samples in the defined sequence length depends on the particular interval/sampling rate. For example, if the interval between samples is 1 minute and the sequence length is 15 minutes, there are 15 samples inside the input data fed to the statistical models. The prediction length represents the scheduled time after the current traffic when we need to predict the traffic.

Therefore, we have run experiments for 60 parameter combinations ($3 \times 4 \times 5$) with HA, ARIMA, SVR, and SVR-Graph. We have calculated the average RMSE and MAE across all these techniques to sort out ten best performing combinations. Table \ref{performanceAll} shows the detailed RMSE, MAE, and CORR for these eight best performing combinations. We have arranged them according to the sample interval 0.5 minute and 1.0 minute in the table \ref{performanceAll}.

From table \ref{performanceAll}, we observe that sampling interval 0.5 minute and interval 1.0 minute had yielded the best results among all combinations where prediction lengths are 5, 15, 30, and 45 minutes.  Specifically, the combination 0.5 min/60 min/5 min achieved the best performance with the Historical Average (HA) model: 73.13 (RMSE), 42.89 (MAE), and 0.64 (CORR). On the other hand, 1.0 min/60 min/5 min combination with ARIMA yielded the second-best result: 73.26 (RMSE), 42.70 (MAE), and 0.64 (CORR). So, please note that these two combinations' performances are very close to each other.

While analyzing these two combinations' performances, we observed that the RMSE and MAE are large, whereas the CORR is very good. High CORR indicates that the trends between the ground truth and predicted traffic have matched with each other with a high positive correlation. To investigate the high RMSE and MAE, we have analyzed the performance for each intersection. For 0.5 min/60 min/5 min combination, we observed that two intersections (Kazi para overbridge and  Senpara Parbata) have unusually high RMSE (201.92 and 126.88 ) and MAE (159. 32 and 88.58) for HA. If we disregard these two intersections, the average RMSE decreases significantly from 73.13 to 39.81, and the average MAE gets down from 42.70 to 35.34. The same situation is also true for 1-min/60-min/5-min combination, where Kazi para overbridge has RMSE= 202.37 and MAE = 159.52 for ARIMA. The numbers for Senpara Parbata are:  RMSE= 127.07 and MAE = 88.74.  After discarding these two intersections, the RMSE decreased from 73.26 to 39.81 and the MAE from 43.02 to 35.07. Therefore, the traffic pattern for these intersections might not be captured by HA and ARIMA, respectively.

Now, we run experiments to see the difference in traffic patterns between weekdays and weekends.  To do that, we have run experiments using the following four different train and test data with the best combination from table \ref{performanceAll}.

\begin{enumerate}
    \item training with traffic data for 14 weekdays, testing with data for six weekdays.  
    \item training with data for 20 weekdays, testing with ten weekend days.
    \item training with data for seven weekend days, testing with data for three weekdays.  
    \item training with data for seven weekend days, testing with three weekend days.
\end{enumerate}

From the table \ref{performanceWeekdays}, it is easy to note that the $4^{th}$ experiment has yielded the best result when both the train and test sets are from weekend data. HA and ARIMA models got very close performance scores in predicting the traffic. Therefore, we can conclude that traffic patterns on weekends are similar enough to be predicted by HA and ARIMA models.

In contrast, the $1^{st}$ experimental result suggests that the weekdays' traffic patterns are very much dynamic and intermittent that the HA and ARIMA cannot capture them very well. The same result has been observed for $3^{rd}$ experiments where the HA and ARIMA models have produced erroneous predictions for weekdays while they have been trained with data weekends. The result for the $2^{nd}$ experiments lies in-between.

We can conclude from these experiments that the traffic pattern for weekends can be modeled with simple statistical and machine learning models. However, to learn the weekdays' traffic pattern with large variabilities, we need more sophisticated machine learning algorithms such as graph convolutional networks or recurrent neural networks. Since this paper focuses on the data collection, we plan to perform further analysis using graphical models in the future. With the proposed technique, we can collect data for the entire city in an automated manner. We also plan to incorporate a longer time frame to identify the traffic patterns for the weekday variations.

\begin{table}[]
\caption{Results for weekdays and weekend experiments with 0.5 minute sampling interval, 60 minute sequence length, and 5 minute prediction length}
\label{performanceWeekdays}
\begin{adjustbox}{width=\textwidth}
\begin{tabular}{|c|c|c|c|c|c|c|c|}

\hline
Train/Test   data                                                                                        & \begin{tabular}[c]{@{}c@{}}Sample Interval\end{tabular} & \begin{tabular}[c]{@{}c@{}}Sequence Length\end{tabular} & \begin{tabular}[c]{@{}c@{}}Predicted Length \end{tabular} & Model      & RMSE           & MAE            & CORR           \\ \hline
\multirow{4}{*}{\begin{tabular}[c]{@{}c@{}}Train -14- weekdays \\ Test -6- weekdays\end{tabular}}        & \multirow{4}{*}{0.5min}                                       & \multirow{4}{*}{60min}                                          & \multirow{4}{*}{5min}                                             & HA         & 78.05          & 47.86          & 0.65          \\ \cline{5-8} 
                                                                                                         &                                                               &                                                                 &                                                                   & SVR        & 91.83          & 67.37          & 0.58          \\ \cline{5-8} 
                                                                                                         &                                                               &                                                                 &                                                                   & SVR\_GRAPH & 92.47          & 68.55          & 0.57          \\ \cline{5-8} 
                                                                                                         &                                                               &                                                                 &                                                                   & ARIMA      & 78.09          & 47.85          & 0.65          \\ \hline
\multirow{4}{*}{\begin{tabular}[c]{@{}c@{}}Train  -20- weekdays \\  Test -10- weekend days\end{tabular}} & \multirow{4}{*}{0.5min}                                       & \multirow{4}{*}{60min}                                          & \multirow{4}{*}{5min}                                             & HA         & 72.82          & 42.15          & 0.63          \\ \cline{5-8} 
                                                                                                         &                                                               &                                                                 &                                                                   & SVR        & 97.17          & 72.35          & 0.52          \\ \cline{5-8} 
                                                                                                         &                                                               &                                                                 &                                                                   & SVR\_GRAPH & 95.41          & 71.74          & 0.53          \\ \cline{5-8} 
                                                                                                         &                                                               &                                                                 &                                                                   & ARIMA      & 72.85          & 42.07          & 0.63          \\ \hline
\multirow{4}{*}{\begin{tabular}[c]{@{}c@{}}Train -7- weekend days\\ Test -3- weekdays\end{tabular}}      & \multirow{4}{*}{0.5min}                                       & \multirow{4}{*}{60min}                                          & \multirow{4}{*}{5min}                                             & HA         & 78.23          & 48.48          & 0.64          \\ \cline{5-8} 
                                                                                                         &                                                               &                                                                 &                                                                   & SVR        & 85.12          & 58.57          & 0.58          \\ \cline{5-8} 
                                                                                                         &                                                               &                                                                 &                                                                   & SVR\_GRAPH & 85.44          & 59.16          & 0.56          \\ \cline{5-8} 
                                                                                                         &                                                               &                                                                 &                                                                   & ARIMA      & 78.35          & 48.56          & 0.64          \\ \hline
\multirow{4}{*}{\begin{tabular}[c]{@{}c@{}}Train -7- weekend days\\ Test -3- weekend days\end{tabular}}  & \multirow{4}{*}{0.5min}                                       & \multirow{4}{*}{60min}                                          & \multirow{4}{*}{5min}                                             & HA         & \underline{\textbf{62.68}} & \underline{\textbf{32.03}} & \underline{\textbf{0.57}} \\ \cline{5-8} 
                                                                                                         &                                                               &                                                                 &                                                                   & SVR        & 81.70          & 55.54          & 0.44          \\ \cline{5-8} 
                                                                                                         &                                                               &                                                                 &                                                                   & SVR\_GRAPH & 78.81          & 53.77          & 0.44          \\ \cline{5-8} 
                                                                                                         &                                                               &                                                                 &                                                                   & ARIMA      & \underline{\textbf{62.68} }         & \underline{\textbf{32.02}}          & \underline{\textbf{0.57}}          \\ \hline
\end{tabular}
\end{adjustbox}
\end{table}

Similar results have been observed for combination 1 min/60 min/ 5 min combination as shown in Table \ref{performanceWeekdays2}.
\begin{table}[]
\caption{Additional results for weekdays and weekend experiments with 1 minute sample interval, 60 minute sequence length, and 5 minute prediction length}
\label{performanceWeekdays2}
\begin{adjustbox}{width=\textwidth}
\begin{tabular}{|c|c|c|c|c|c|c|c|}
\hline
Train/Test   data                                                                                      & Sampling Rate           & Sequence Length        & Predicted   Length    & Model      & RMSE  & MAE   & CORR \\ \hline
\multirow{4}{*}{\begin{tabular}[c]{@{}c@{}}Train -14- weekdays\\ Test - 6 -weekdays\end{tabular}}      & \multirow{4}{*}{1.0min} & \multirow{4}{*}{60min} & \multirow{4}{*}{5min} & HA         & 78.04 & 47.58 & 0.65 \\ \cline{5-8} 
                                                                                                       &                         &                        &                       & SVR        & 91.30 & 66.82 & 0.58 \\ \cline{5-8} 
                                                                                                       &                         &                        &                       & SVR\_GRAPH & 90.76 & 66.88 & 0.58 \\ \cline{5-8} 
                                                                                                       &                         &                        &                       & ARIMA      & 77.66 & 47.24 & 0.66 \\ \hline
\multirow{4}{*}{\begin{tabular}[c]{@{}c@{}}Train -20- weekdays\\ Test -10 - weekend days\end{tabular}} & \multirow{4}{*}{1.0min} & \multirow{4}{*}{60min} & \multirow{4}{*}{5min} & HA         & 73.03 & 42.36 & 0.64 \\ \cline{5-8} 
                                                                                                       &                         &                        &                       & SVR        & 97.16 & 72.32 & 0.52 \\ \cline{5-8} 
                                                                                                       &                         &                        &                       & SVR\_GRAPH & 94.00 & 70.22 & 0.53 \\ \cline{5-8} 
                                                                                                       &                         &                        &                       & ARIMA      & 72.69 & 42.01 & 0.64 \\ \hline
\multirow{4}{*}{\begin{tabular}[c]{@{}c@{}}Train -7- weekend days \\  Test -3- weekdays\end{tabular}}  & \multirow{4}{*}{1.0min} & \multirow{4}{*}{60min} & \multirow{4}{*}{5min} & HA         & 78.23 & 48.23 & 0.65 \\ \cline{5-8} 
                                                                                                       &                         &                        &                       & SVR        & 84.81 & 58.09 & 0.58 \\ \cline{5-8} 
                                                                                                       &                         &                        &                       & SVR\_GRAPH & 84.52 & 58.02 & 0.57 \\ \cline{5-8} 
                                                                                                       &                         &                        &                       & ARIMA      & 77.97 & 47.98 & 0.65 \\ \hline
\multirow{4}{*}{\begin{tabular}[c]{@{}c@{}}Train -7-weekend days\\ Test-3-weekend days\end{tabular}}   & \multirow{4}{*}{1.0min} & \multirow{4}{*}{60min} & \multirow{4}{*}{5min} & HA         & \underline{\textbf{63.62}} & \underline{\textbf{32.82}} & \underline{\textbf{0.57}} \\ \cline{5-8} 
                                                                                                       &                         &                        &                       & SVR        & 82.63 & 55.85 & 0.43 \\ \cline{5-8} 
                                                                                                       &                         &                        &                       & SVR\_GRAPH & 78.55 & 52.99 & 0.44 \\ \cline{5-8} 
                                                                                                       &                         &                        &                       & ARIMA      & \underline{\textbf{63.44}} & \underline{\textbf{32.69}} & \underline{\textbf{0.57}} \\ \hline
\end{tabular}
\end{adjustbox}
\end{table}

\section{Conclusion}
\label{conclusion}
This paper proposes a novel approach for traffic data collection in developing countries with minimal cost and human effort. We have proposed to use traffic information from Google Map Service in a novel way so that the traffic data for any road segments/intersections can be collected seamlessly from any location in a city. Therefore, the proposed approach is scalable from a small area to the entire metropolitan city. Further, we have implemented some statistical and machine learning models: historical averages (HA), Support Vector Regression (SVR), Support Vector Regression with Graph (SVR-Graph), and Auto-Regressive Integrated Moving Average (ARIMA) on the traffic data of November 2019 at Mirpur, Dhaka Bangladesh. The experimental results suggest that these simple models can effectively model the traffic congestion of Mirpur, Dhaka. Also, the results show the efficacy of the collected traffic data in forecasting future congestion. 

In the future, we plan to incorporate more intersections to develop a comprehensive dataset of the city road network. As mentioned in the discussion, we will apply more sophisticated machine learning models such as Graph Convolution Network and Recurrent Neural Networks on the data to accurately estimate the traffic pattern. 
\section{ACKNOWLEDGMENT}
This work is partially funded by the Independent University, Bangladesh, and the ICT Division of ICT Ministry, Bangladesh.
%
%

\end{document}